\documentstyle[12pt,epsfig]{article}
\begin{document}
\noindent
\begin{center}
{\Large {\bf Coincidence Problem
in Holographic $f(R)$ Gravity }}\\ \vspace{2cm}
 ${\bf Yousef~Bisabr}$\footnote{e-mail:~y-bisabr@srttu.edu.}\\
\vspace{.5cm} {\small{Department of Physics, Shahid Rajaee Teacher
Training University,
Lavizan, Tehran 16788, Iran}}\\
\end{center}
\vspace{1cm}
\begin{abstract}
It is well-known that $f(R)$ gravity models formulated in Einstein conformal frame
are equivalent to Einstein gravity together with a minimally
coupled scalar field. In this case, the scalar field couples with the matter sector and the coupling term is given by the conformal factor.
We apply the holographic principle to such interacting models.  In a spatially flat universe, we
show that the Einstein frame representation of $f(R)$ models leads to a constant ratio of energy densities of dark matter to dark energy.

\end{abstract}
\vspace{3cm}
\section{Introduction}
It is strongly believed that our universe is in a phase
experiencing an accelerated expansion.  The simplest candidate to produce this cosmic
speed-up is the cosmological constant, the energy density
associated with quantum vacuum.  However, there are several problems for associating cosmic acceleration with the cosmological
constant.  First, theoretical estimates on its value are many
order of magnitude larger than observations \cite{win}.  Second,
it is simply a constant, namely that it is not diluted with
expansion of the universe.  This latter is specifically important
in the sense that there are observational evidence \cite{R}
demonstrating that the cosmic acceleration is a recent phenomena and
the universe must have passed through a deceleration phase in the
early stages of its evolution.  This deceleration phase is
important for successful nucleosynthesis as well as for the
structure formation. We therefore need a field evolving during
expansion of the universe in such a way that its dynamics makes
the deceleration parameter have a signature flip from  positive in
the early stages of matter dominated era to negative in the
present
stage \cite{b}.  There is also another
problem which is the focus of the present note.  It concerns with the coincidence between the observed vacuum
energy density and the current matter density.  While these two
energy components evolve differently as the universe expands,
their contributions to total energy density of the universe in the
present epoch are the same order of magnitude.  \\
As a different point of view, cosmic acceleration may be interpreted as evidence either for
existence of some exotic matter components or for modification of
the gravitational theory.  In the first route of interpretation
one can take a mysterious cosmic fluid with
sufficiently large and negative pressure, dubbed dark energy.
These models are usually invoked a scalar field which during its
evolution takes negative pressure by rolling down a proper
potential.  In the second route, however, one attributes the
accelerating expansion to a modification of general relativity. A
particular class of models that has recently drawn a significant
amount of attention is the so-called $f(R)$ gravity models
\cite{r}. These models propose a modification of
Einstein-Hilbert action so that the scalar curvature is replaced
by some arbitrary function $f(R)$. \\
Recently, different models inspired by holographic principle have been proposed to explain the cosmic acceleration.
The basic
idea is that the number of degrees
of freedom of a physical system scales with its bounding area
rather than with its volume \cite{suss}.  For an effective quantum field
theory in a box of size $L$ with an ultraviolet (UV) cutoff
$\Lambda$, the entropy $S$ scales extensively as $S\sim
L^{3}\Lambda^{3}$. However, the peculiar thermodynamics of black
holes has led Bekenstein \cite{bek} to postulate that the maximum
entropy in a box of volume $L^{3}$ behaves non-extensively,
growing as the area of the box.  In this sense there is a
so-called Bekenstein entropy bound
\begin{equation}
S=L^{3}\Lambda^{3}\leq S_{BH}\equiv \pi L^{2}M^{2}_{p}
\label{0a}\end{equation} where $S_{BH}$ is the entropy of a black
hole of radius $L$, and $M_{p}\equiv (8\pi G)^{-\frac{1}{2}}$
stands for the reduced Planck mass.  It is important that in this
relation the length scale $L$ providing an Infrared (IR) cutoff is
determined by the UV cutoff $\Lambda$ and can not be chosen
independently. However, such a non-extensive scaling law seems to
provide a breakdown of quantum field theory at large scales.  To
reconcile this breakdown with the success of local quantum field
theory in describing observed particle phenomenology, Cohen et al.
\cite{co} proposed a more restrictive bound.  Since the maximal
energy density in the effective theory is of the order
$\rho_{\Lambda}=\Lambda^{4}$, requiring that the energy in a given
volume not to exceed the energy of a black hole of the same size
results in the constraint
\begin{equation}
L^{3}\rho_{\Lambda}\leq L M^{2}_{p} \label{02}\end{equation} If we
take the largest value of the length scale $L$ as the IR cutoff
saturating the inequality (\ref{02}), we then obtain the
holographic dark energy density
\begin{equation}
\rho_{\Lambda}=3c^{2} M^{2}_{p}L^{-2} \label{03}\end{equation} in
which $3c^{2}$ is a numerical constant.  It is interesting to note
that if the length scale $L$ is characterized by the size of the
universe, the Hubble scale $H^{-1}$, then equation (\ref{03})
gives a vacuum energy density of the right order of magnitude
consistent with observations \cite{co}.  It is, however, pointed out
that this yields a wrong equation of state parameter for dark
energy, and other possible values for $L$ should be chosen such as
the size of the future event horizon \cite{li} \cite{cite}.  This conclusion
is, however,  based on the assumption that energy densities of dark energy and dark matter
evolve independently.  It is shown
\cite{pp} that, if there is \emph{any} interaction between these two
components the identification of $L$ with $H^{-1}$ is possible.  In particular, the authors of \cite{pp} argued that such an
identification
necessarily implies a constant ratio of the energy densities of the two components \emph{regardless of the details of the interaction}. \\
In the present note, we investigate the coincidence problem in the context of holographic  $f(R)$ gravity models\footnote{Holographic principle
has been already applied to different modified gravity models. See, for instance, \cite{h}.}.
In $f(R)$ models the dynamical variable of the vacuum sector
is the metric tensor and the corresponding field equations are fourth order.  This
dynamical variable can be replaced
by a new pair which consists of a conformally rescaled metric and a scalar partner.  Moreover, in terms of the new set
of variables the field equations are those of General Relativity.  The original set of variables
is commonly called Jordan conformal frame and the transformed set whose dynamics is described by Einstein field equations
is called Einstein conformal frame. The dynamical
equivalence of Jordan and Einstein conformal frames does not
generally imply that they are also physically equivalent.  In fact,
it is shown that some physical systems can be differently
interpreted in different conformal frames \cite {soko} \cite{no}.
The physical status of the two conformal frames is an open
question which we are not going to address here.\\
We will work in Einstein conformal frame.  The motivation is that in this frame there is a coupling between the scalar degree of freedom and matter
sector induced by the conformal transformation.  In this context, we have already studied the coincidence problem without any use of holographic principle \cite{bi}.  We have shown that
the requirement of a constant ratio of energy densities of the two components, puts some constraints on the functional form of the $f(R)$ function.  Here we apply the holographic principle to dark energy
density corresponding to the scalar degree of freedom.
The IR cutoff is identified with the Hubble scale. We shall show that this interacting holographic dark energy leads to a stationary ratio of energy densities corresponding to dark energy and matter sector in a spatially flat universe regardless
of the details of the interaction term.  The
distinguished feature of the present work is that the interaction term is given by a particular configuration of the $f(R)$ function.  We use
this fact to argue that Einstein frame representation of $any$ holographic $f(R)$ model may address the coincidence problem in a spatially flat universe.
\section{The Model}
Let us start with introducing the action for an $f(R)$
gravity theory in the Jordan frame
\begin{equation}
S_{JF}= \frac{1}{2}\int d^{4}x \sqrt{-g}~M_p^2~ f(R) +S_{m}(g_{\mu\nu}, \psi)\label{b1}\end{equation}
where $g$ is the
determinant of $g_{\mu\nu}$ and $S_{m}$ is the action
of (dark) matter which depends on the metric $g_{\mu\nu}$ and some (dark) matter
field $\psi$.  Stability in matter sector (the Dolgov-Kawasaki
instability \cite{dk}) imposes some conditions on the functional
form of $f(R)$ models.  These
conditions require that the first and the second derivatives of
$f(R)$ function with respect to the Ricci scalar $R$ should be
positive definite.  The positivity of the first derivative ensures
that the scalar degree of freedom is not tachyonic and positivity
of the second derivative tells us that graviton is not a ghost.\\
It is well-known that $f(R)$ models are equivalent to models in which
a scalar field minimally couples to gravity with an appropriate
potential function.  In fact, we may use a new set of variables
\begin{equation}
\bar{g}_{\mu\nu} =\Omega~ g_{\mu\nu} \label{b2}\end{equation}
\begin{equation} \phi = \frac{M_p}{2\beta } \ln \Omega
\label{b3}\end{equation}
 where
$\Omega\equiv\frac{df}{dR}=f^{'}(R)$ and
$\beta=\sqrt{\frac{1}{6}}$. This is indeed a conformal
transformation which transforms the above action in the Jordan
frame to the following action in the Einstein frame \cite{soko} \cite{w}
\begin{equation}
S_{EF}=\frac{1}{2} \int d^{4}x \sqrt{-\bar{g}}~\{\frac{1}{M_p^2}
\bar{R}-\bar{g}^{\mu\nu} \nabla_{\mu} \phi~ \nabla_{\nu} \phi
-2V(\phi)\}+ S_{m}(\bar{g}_{\mu\nu}e^{2\beta \phi/M_p}
, \psi) \label{b4}\end{equation}All indices are raised and lowered by $\bar{g}_{\mu\nu}$.  In the Einstein frame, $\phi$ is a minimally
coupled scalar field with a self-interacting potential which is
given by
\begin{equation}
V(\phi(R))=\frac{M_p^2(Rf'(R)-f(R))}{2f'^2(R)}
\label{b5}\end{equation} Note that the conformal transformation
induces the coupling of the scalar field $\phi$ with the matter
sector. The strength of this coupling $\beta$, is fixed to be
$\sqrt{\frac{1}{6}}$ and is the same for all types of matter
fields.  In the action (\ref{b4}), we take $\bar{g}^{\mu\nu}$ and
$\phi$ as two independent field variables and variations of the
action yield the corresponding dynamical field equations.
Variation with respect to the metric tensor $\bar{g}^{\mu\nu}$,
leads to
\begin{equation}
\bar{G}_{\mu\nu}=M_p^{-2}~(\bar{T}^{\phi}_{\mu\nu}+
\bar{T}^{m}_{\mu\nu}) \label{b6}
\label{b7}\end{equation} where
\begin{equation}
\bar{T}^{\phi}_{\mu\nu}=\nabla_{\mu}\phi
\nabla_{\nu}\phi-\frac{1}{2}\bar{g}_{\mu\nu}\nabla^{\gamma}\phi
\nabla_{\gamma}\phi-V(\phi)\bar{g}_{\mu\nu}
\label{b8}\end{equation}
\begin{equation}
\bar{T}^m_{\mu\nu}=\frac{-2}{\sqrt{-\bar{g}}}\frac{\delta S_{m}(\bar{g}_{\mu\nu}, \psi)}{\delta \bar{g}^{\mu\nu}} \label{b9}\end{equation} are
stress-tensors of the scalar field and the matter field system.
The trace of (\ref{b7}) is
\begin{equation}
\nabla^{\gamma}\phi
\nabla_{\gamma}\phi+4V(\phi)-M_p^2~\bar{R}=\bar{T}^m\label{b8-1}\end{equation}
which differentially relates the trace of the matter stress-tensor $\bar{T}^{m}=\bar{g}^{\mu\nu}\bar{T}^m_{\mu\nu}$
to $\bar{R}$.  Variation of the action
(\ref{b4}) with respect to the scalar field $\phi$, gives
\begin{equation}
\bar{\Box}\phi-\frac{dV(\phi)}{d\phi}=-\frac{\beta}{M_p} \bar{T}^{m}
\label{b11}\end{equation}
It is important to note that the two
stress-tensors $\bar{T}^m_{\mu\nu}$ and $\bar{T}^{\phi}_{\mu\nu}$
are not separately conserved.
Instead, they satisfy the following equations
\begin{equation}
\bar{\nabla}^{\mu}\bar{T}^{m}_{\mu\nu}=-\bar{\nabla}^{\mu}\bar{T}^{\phi}_{\mu\nu}= \frac{\beta}{M_p} \nabla_{\nu}\phi~\bar{T}^{m}\label{b13}\end{equation} We apply the field equations in a
spatially flat homogeneous and isotropic cosmology described by
Friedmann-Robertson-Walker spacetime
\begin{equation}
ds^2=-dt^2+a^2(t)(dx^2+dy^2+dz^2)
\end{equation}
where $a(t)$ is the scale factor. To do this, we take
$\bar{T}^m_{\mu\nu}$ and $\bar{T}^{\phi}_{\mu\nu}$ as the stress-tensors of a pressureless perfect fluid with energy density
$\bar{\rho}_{m}$, and a perfect fluid with energy density
$\rho_{\phi}=\frac{1}{2}\dot{\phi}^2+V(\phi)$ and pressure
$p_{\phi}=\frac{1}{2}\dot{\phi}^2-V(\phi)$, respectively. In this
case, (\ref{b7}) and (\ref{b11}) take the form \footnote{Hereafter we will use unbarred characters in the Einstein frame.}
\begin{equation}
3H^2=M_p^{-2}(\rho_{\phi}+\rho_{m})
\label{b14}\end{equation}
\begin{equation}
2\dot{H}+3H^2=-M_p^{-2}\omega_{\phi}\rho_{\phi}
\label{b14-1}\end{equation}
\begin{equation}
\ddot{\phi}+3H\dot{\phi}+\frac{dV(\phi)}{d\phi}=-\frac{\beta}{M_p} \rho_{m}
\label{b15}\end{equation} where
$\omega_{\phi}=\frac{p_{\phi}}{\rho_{\phi}}$ is equation of state parameter of the scalar field $\phi$, and overdot indicates differentiation with respect
to cosmic time $t$.  The trace equation (\ref{b8-1}) and the conservation equations
(\ref{b13}) give, respectively,
\begin{equation}
\dot{\phi}^2+M_p^2~R-4V(\phi)=\rho_{m}
\label{b16}\end{equation}
\begin{equation}
\dot{\rho}_{m}+3H\rho_{m}=Q \label{b17}\end{equation}
\begin{equation}
\dot{\rho}_{\phi}+3H(\omega_{\phi}+1)\rho_{\phi}=-Q
\label{b18}\end{equation} where
\begin{equation}
Q=\frac{\beta}{M_p} \dot{\phi}\rho_{m}
\label{b-18}\end{equation} is the interaction term.  This term
vanishes only for $\phi$~=~const., which due to (\ref{b3}) it happens
when $f(R)$ linearly depends on $R$. The direction of energy
transfer depends on the sign of $Q$ or $\dot{\phi}$.  For
$\dot{\phi}>0$, the energy transfer is from dark energy to dark
matter and for $\dot{\phi}<0$ the reverse is true\footnote{Dark energy and dark matter are the most important energy/mass components contained
in the universe.  However, there is no experiment to show that these components interact with ordinary matter systems.  It is quite possible that these components interact with each other while not being coupled to standard model particles.}.\\
Let us consider time evolution of the ratio $r\equiv \rho_{m}/\rho_{\phi}$ ,
\begin{equation}
\dot{r}=\frac{\dot{\rho}_{m}}{\rho_{\phi}}-r\frac{\dot{\rho}_{\phi}}{\rho_{\phi}}
\label{c1}\end{equation}  If we combine the latter with the
balance equations (\ref{b17}) and (\ref{b18}), we obtain
\begin{equation}
\dot{r}=3Hr
[\omega_{\phi}+(1+\frac{1}{r})\frac{\Gamma}{3H}]
\label{c2}\end{equation} where
\begin{equation}
\Gamma=\frac{Q}{\rho_{\phi}}=\frac{\beta}{M_p}r\dot{\phi}
\label{c2c}\end{equation}
is the decay rate.  Now we apply the holographic relation to dark energy density $\rho_{\phi}$ with $L=H^{-1}$,
\begin{equation}
\rho_{\phi}=3c^2 M_p^2 H^2 \label{c3}\end{equation}where $c^2$ is a numerical constant introduced for convenience.  This gives
\begin{equation}
\dot{\rho}_{\phi}=6c^2 M_p^2 H \dot{H}
\label{c4}\end{equation}
We combine (\ref{c3}) with (\ref{b14-1}) to obtain
\begin{equation}
\dot{H}=-\frac{3}{2}H^2(1+\frac{\omega_{\phi}}{r+1})(r+1)c^2
\label{c55}\end{equation}
One can easily check that
\begin{equation}
c^2=\frac{1}{r+1}
\label{c5c}\end{equation}
which reduces (\ref{c55}) to
\begin{equation}
\dot{H}=-\frac{3}{2}H^2(1+\frac{\omega_{\phi}}{r+1})
\label{c5}\end{equation}
Substituting this into (\ref{c4}) gives
\begin{equation}
\dot{\rho}_{\phi}=-9c^2 M_p^2 H^3 (1+\frac{\omega_{\phi}}{r+1})
\label{c6}\end{equation}
When we put the latter together with the holographic relation (\ref{c3}) into the balance equation (\ref{b18}), we obtain
\begin{equation}
\omega_{\phi}=-(1+\frac{1}{r})\frac{\Gamma}{3H}
\label{c7-1}\end{equation}
This yields the equation of state parameter in terms of $r$ and the decay rate $\Gamma$.  Note that there is no non-interacting
limit in our case since $\Gamma=0$ corresponds to $\phi$~=~const., or equivalently, $\Lambda$CDM model. \\From the expression (\ref{c7-1}), it is clear that when $\frac{\Gamma}{3H}<<1$ the equation of state of dark energy is closely related to that of the dust.  This can also be seen from the balance equation (\ref{b18}).  In the other limiting case, when $\frac{\Gamma}{3H}>>1$ one takes $\omega_{\phi}<<-1$.  This behavior correspond to a signature flip of the deceleration parameter.  To see this, we write the
deceleration parameter as,
\begin{equation}
q=-1-\frac{\dot{H}}{H^2}
\end{equation}
This relation together with (\ref{c5}) and (\ref{c7-1}) results in
\begin{equation}
q=\frac{1}{2}(1-\frac{\Gamma}{rH})
\label{qq}\end{equation}
which in the above two limiting cases changes the sign from $q>0$ to $q<0$, respectively.\\
As our main observation, we remark that if one uses (\ref{c7-1}) in
the relation (\ref{c2}) one then takes $\dot{r}=0$ or $r=$~constant.  The reasoning is simple : from the holographic
relation (\ref{c3}) one infers that $\rho_{\phi}$  scales like the critical density $\rho_c=3M_p^2H^2$.  As a consequence, the density parameter corresponding to $\phi$ must be a constant so that $\Omega_{\phi}=\frac{\rho_{\phi}}{\rho_c}=c^2$.  With this result, the Friedmann equation $\Omega_{\phi}+\Omega_m=1$ results in $\Omega_m=1-c^2$. Thus $\rho_m$ has the same scaling
as $\rho_{\phi}$ and the ratio $r$ is a constant.  There are some remarks to do with respect to this result.  First, it is independent
of the details of the decay rate $\Gamma$ or the interaction $Q$.  Since the interaction is given by the shape of the $f(R)$ function
we conclude that applying holographic principle to the dark energy density $\rho_{\phi}=3c^2 M_p^2 H^2$ necessarily leads to a constant
ratio of energy densities $r=\rho_m/\rho_{\phi}$, irrespective of the form of the $f(R)$ function\footnote{There are
different constraints on the configuration of a viable $f(R)$ function, such as constraints coming from Dolgov-Kawasaki
instability issue \cite{dk} or constraints related to local gravity experiments \cite{bb}. However, resolution of the coincidence problem
 in our analysis does not put any constraint on the form of the $f(R)$ function. }.  Second, it is the consequence of
the identification $L=H^{-1}$.  If one takes other length scales such as
event horizon $L_{e}=a(t)\int_{t}^{\infty}\frac{dt'}{a(t')}$ or particle horizon $L_p=a(t)\int_{0}^{t}\frac{dt'}{a(t')}$
as the IR cutoff, then scaling of $\rho_{\phi}$ will be different from that of the critical density and the ratio $r$ is no longer stationary.
In contrary to the ratio $r$, accelerating expansion requires
particular configurations of $f(R)$ functions.  This is clear from the expression (\ref{qq}) which the requirement that $q<0$ automatically
sets a constraint on the decay rate.\\
It is quite possible that the constancy of the ratio of energy densities $r$ is a recent phenomenon.  The two energy components can evolve
differently as the universe expands until the present epoch which their contributions to total energy density takes a constant configuration
and of the same order of magnitude.  To model this behavior, we assume that during evolution of the universe the holographic
relation holds as the unsaturated form $\rho_{\phi}\leq M_p^2 H^2$ and the saturated relation (\ref{c3}) is a recent phenomenon.  This is equivalent to
assume that
\begin{equation}
\rho_{\phi}=3\alpha(t)M_p^2 H^2
\label{ab}\end{equation}
where $\alpha(t)\leq c^2$ with $\alpha(t)$ being a parameter which evolves with cosmic expansion.  Note that the IR cutoff does not
change and remains $L=H^{-1}$.  In fact, variation of $\alpha(t)$ in the relation (\ref{ab}) characterizes the degree of the saturation
in the holographic bound $\rho_{\phi}\leq M_p^2 H^2$.  \\In this case, the relation (\ref{c7-1}) takes the form
\begin{equation}
\omega_{\phi}=-(1+\frac{1}{r})(\frac{\Gamma}{3H}+\frac{\dot{\alpha}}{3H\alpha})
\label{c7-11}\end{equation}
We may combine the latter with (\ref{c2}) to obtain
\begin{equation}
\frac{\dot{\alpha}}{\alpha}=-\frac{\dot{r}}{r+1}
\label{bb}\end{equation}
This has a solution $\alpha(t)=\frac{1}{r+1}$ which is compatible with (\ref{c5c}) up to an integration constant.  Since $\dot{\alpha}>0$ by construction, $r$ should be a decreasing function of time.  This behavior allows $\omega_{\phi}$ in (\ref{c7-11}) to be more negative compared with the case that $\alpha$=constant.
\section{Conclusion}
We have considered the Einstein frame representation of a general $f(R)$ gravity model and apply the holographic relation
to the energy density $\rho_{\phi}$ corresponding to the scalar degree of freedom of the metric tensor.  Taking the Hubble radius as the IR
cutoff, we observe that the ratio $r=\rho_m/\rho_{\phi}$ takes a constant configuration for any $f(R)$ function
in a spatially flat universe.  It should be noted that the choice $L=H^{-1}$ attributes an energy density to $\rho_{\phi}$ consistent with observations \cite{co}.  Thus the two different features of the cosmological constant problem, namely the fine tuning and the cosmic coincidence problems, may be addressed in this context.\\

\end{document}